\documentclass[prl,twoside,twocolumn,showpacs,aps,longbibliography]{revtex4-1}
\usepackage{amsmath}
\usepackage{amssymb}
\usepackage{amsfonts}
\newcommand{\vecg}{\boldsymbol}
\renewcommand{\vec}{\textbf}

\DeclareMathOperator{\tr}{Tr} 
\usepackage{subfig}
\usepackage{graphicx}

\begin{document}

\title{Nonlinear evolution and signaling}

\author{Jakub Rembieli\'nski}
\email{jaremb@uni.lodz.pl}
\author{Pawe{\l}{} Caban}
\email{P.Caban@merlin.phys.uni.lodz.pl}

\affiliation{Department of Theoretical Physics,\\
	Faculty of Physics and Applied Informatics, University of Lodz\\
Pomorska 149/153, 90-236 {\L}{\'o}d{\'z}, Poland}
\date{\today}

\begin{abstract}
We propose a condition, called convex quasi-linearity, for deterministic 
nonlinear quantum evolutions. Evolutions satisfying this condition
do not allow for arbitrary fast signaling, therefore, they cannot 
be ruled out by a standard argument.
We also give an explicit example of a nonlinear qubit evolution
satisfying quasi-linearity.
\end{abstract}
\maketitle

\textit{Introduction.}---From almost of century, quantum mechanics (QM), 
in his version based
on the Hilbert space, was formulated as a linear theory preserving the
superposition principle for pure states. 
However, many authors, for different reasons, undertaken attempts to
generalize this theory by including nonlinear operations, too
(see, e,g, \cite{BM1976_Nonlinear,Weinberg1989_Nonlin_Annals,Weinberg1989_PRL_nonlin,DG1992_Nonlin_PLA,DG1996_Nonlin_PhysRevA.54.3764} and references
therein).
In fact a part of the QM formalism related to the measurement description
uses nonlinear (stochastic) operations such as the selective projection
postulate.  
Most of proposed attempts lie in replacement of the linear
time evolution of quantum systems by a nonlinear one
(see, e.g., \cite{BM1976_Nonlinear,Weinberg1989_PRL_nonlin}).
It is widely believed that such deterministic nonlinear generalizations
of the Schr\"odinger equation allow for signaling, i.e., allows one to send
signals over arbitrarily large distances in a finite time (see, e.g., \cite{GR1995_relevant_Sch}).
Arguments supporting this claim (in the context of the Weinberg model
\cite{Weinberg1989_PRL_nonlin}) were clearly given by Gisin in 
Ref.~\cite{Gisin1990_Weinberg_nonlin}, compare also
\cite{Polchinski1991_on_Weinberg,Czachor1991}.
Gisin's arguments are based on the observation that
deterministic nonlinear time evolution destroys equivalence of quantum
ensembles defining the same mixed state of the considered system. 
As a consequence, it creates a possibility of an instantaneous communication
for space-like separated observers with the help of systems of 
entangled particles. 
Evidently such a possibility is in an apparent conflict with the special relativity.

Let us mention here that some authors gave arguments that 
deterministic nonlinear dynamics in special circumstances does not allow 
for signaling \cite{Polchinski1991_on_Weinberg,Czachor1998_nonlocal-eq-NLQM,CD2002_Correl-exp-nonlin_PLA,Kent2005_Nonlin_superluminality,FSSG2004_NLQM-not-supraluminal,Jordan2009_Why-q-dyn-is-linear,HCh2017_Born_NLQM}.
However, these arguments seems to be insufficient (compare \cite{Mielnik2001_NLQM,Jordan2010_Test_NLQM}).
For example, Czachor and Doebner \cite{CD2002_Correl-exp-nonlin_PLA}
approach implies the modification of the state reduction postulate
while Helou and Chen \cite{HCh2017_Born_NLQM} postulate the
extension of the Born rule.
Our goal is different: we are looking for such an extension of deterministic
linear dynamics that does not allow for faster than light signaling and at
the same time does not require anything to be changed in the remaining
part of quantum formalism. 

Notice also that there exist nonlinear stochastic evolution equations free of the 
problems with signaling \cite{GR1995_relevant_Sch}. 
Such models
were proposed in various contexts, one of the most important are
collapse models (see Ref.~\cite{GRW1986,BLSSU2013} and references therein).

In this Letter we propose a new condition for a deterministic nonlinear
quantum evolutions---quasi-linearity [Eq.~(\ref{quasi_linearity})].
This condition guarantees that evolution preserves the equivalence
of quantum ensembles.
Consequently, the Gisin argument \cite{Gisin1990_Weinberg_nonlin} 
does not work in this  case.

\textit{The Gisin's argument.}---Following Gisin
\cite{Gisin1990_Weinberg_nonlin}, let us 
assume that two distant observers, say A and B, want to establish
instantaneous communication.
In the half of the distance between them there is a source emitting 
pairs of spin-1/2 particles.  
Initial spin state of the particle is the Bell state.
Particles move along the $z$-axis,
one towards A and the second one towards B. 
The observer A performs polarization procedure of the flux of particles directed  to him.
To this end A measures without selection the projector
$\pi_\varphi = \tfrac{1}{2}(I + \vecg{\zeta}_\varphi \cdot \vecg{\sigma})\otimes I$
with a possibility of changing the polarization angle $\varphi$. 
Here, the polarization vector
\begin{equation}
\vecg{\zeta}_\varphi = (\cos\varphi, \sin\varphi,0).
\label{zeta}
\end{equation}
As effect, the Bell state converts into the mixture
\begin{equation}
\rho_\varphi = \pi_\varphi \rho_{\mathsf{Bell}} \pi_\varphi 
+ (I-\pi_\varphi) \rho_{\mathsf{Bell}} (I-\pi_\varphi)
\equiv \tfrac{1}{2} \rho_1 + \tfrac{1}{2} \rho_2.
\end{equation}
Consequently,  the reduced density matrix  $\rho_B=\tr_A(\rho_\varphi)$ 
accessible to the observer B, has the form of the appropriate ensemble
\begin{equation}
\rho_B = \tfrac{1}{2} \rho_{B_{\varphi}^{1}}
+ \tfrac{1}{2} \rho_{B_{\varphi}^{2}}
= \tfrac{1}{2} I,
\label{rho_B}
\end{equation}
with
\begin{align}
\rho_{B_{\varphi}^{1}} & 
= \tfrac{1}{2} \big(I - (\vecg{\zeta}_\varphi \cdot \vecg{\sigma})\big), \\
\rho_{B_{\varphi}^{2}} & 
= \tfrac{1}{2} \big(I + (\vecg{\zeta}_\varphi \cdot \vecg{\sigma})\big).
\end{align}
The ensemble accessible to the observer B is equivalent to the fully depolarized state. 
Thus, any quantum-mechanical  measurement of the observer B is incapable to 
distinguish between different ensembles of this form for different values of 
the polarization angle $\varphi$. 
Consequently, arbitrary fast communication is excluded by the QM rules.

However, according to the Gisin's Gedanken prescription, 
a selective measurement by the observer B can be prefaced by a 
nonlinear evolution (like the Weinberg nonlinear evolution
\cite{Weinberg1989_PRL_nonlin}) applied to the reduced density 
$\rho_B$ and to the related ensembles.
Gisin shows that  the Weinberg nonlinear time evolution does not respect 
equivalence of ensembles so the evolved ensemble is polarization  dependent 
($\vecg{\zeta}_\varphi$ dependent). 
Consequently, changes of polarizations made by the observer A can be registered 
by the observer B, i.e., instantaneous communication is possible. 
Thus, to avoid contradiction with relativity, nonlinear evolution must be 
ruled out of the QM formalism.

We show that this argument does not work for nonlinear evolutions satisfying the 
quasi-linearity condition.

\textit{The convex quasi-linear map.}---Let us consider ensembles of the form 
\begin{equation}
\lambda \rho_a + (1-\lambda) \rho_b \equiv \rho,
\label{rho_ensemble}
\end{equation}
where $0\le\lambda\le1$ and $\rho_a$, $\rho_b$
belong to the convex set $S$ of density matrices.
A trace preserving map $\Phi$ in $S$ is the convex quasi-linear if for each non-negative
$\lambda\le1$ and for arbitrary density operators $\rho_a$, $\rho_b$ 
there exists such $\bar{\lambda}$, $0\le\bar{\lambda}\le1$ that it holds
\begin{equation}
\Phi[\rho] = 
\Phi[\lambda\rho_a + (1-\lambda)\rho_b] = 
\bar{\lambda} \Phi[\rho_a] + (1-\bar{\lambda}) \Phi[\rho_b].
\label{quasi_linearity}
\end{equation}
This class of quantum operations contains linear maps too. 
The essence of this map is that it transforms convex combinations of density operators
into convex combinations of their images so it preserves the convex 
structure of $S$. 
Moreover, it preserves equivalence of ensembles related to
each fixed density matrix.

Usually, it is assumed that deterministic quantum evolution 
preserves mixtures;
it is realized by requirement of linearity of $\Phi$
 \cite{Gisin1989_stochastic_q_dyn_relativ,Jordan1991_assump-Schroedinger,FSSG2004_NLQM-not-supraluminal,Jordan2006_PhysRevA_Assump-imply-linear,BK2015_Faster-then-light-linear}.
However, in our opinion this assumption 
is too restrictive and can be generalized
to the condition (\ref{quasi_linearity}).
In fact, standard quantum-mechanical
selective measurements belong to the class of quasi-linear
transformations. 
Indeed, let us define the trace-preserving nonlinear  map 
corresponding to a selective measurement:
$\Phi(\rho) = \frac{\Pi \rho \Pi}{\tr(\Pi\rho \Pi)}$,
where $\Pi$ is a projector. 
Applying $\Phi$ to the density matrix (\ref{rho_ensemble}) we get
\begin{align}
\Phi[\rho] & = \frac{\lambda \Pi \rho_a \Pi +
	 (1-\lambda)\Pi\rho_b\Pi}{\tr(\Pi \rho)} \nonumber \\
  & = \lambda \frac{\tr(\Pi\rho_a)}{\tr(\Pi\rho)} \Phi[\rho_a] 
  + (1-\lambda) \frac{\tr(\Pi\rho_b)}{\tr(\Pi\rho)} \Phi[\rho_b].
  \label{measurement_quasi_linear}
\end{align}
Now, Eq.~(\ref{measurement_quasi_linear}) has the form
(\ref{quasi_linearity}) provided that
\begin{equation}
1- \lambda \frac{\tr(\Pi\rho_a)}{\tr(\Pi\rho)}  = 
(1-\lambda) \frac{\tr(\Pi\rho_b)}{\tr(\Pi\rho)}.
\label{measurement_coefficient}
\end{equation}
But using (\ref{rho_ensemble}) one can easily verify that
(\ref{measurement_coefficient}) really holds. 
Therefore, Eq.~(\ref{measurement_quasi_linear}) can be cast
in the form (\ref{quasi_linearity}) with
\begin{equation}
\bar{\lambda} = \lambda \frac{\tr(\Pi \rho_a)}{\tr(\Pi \rho)}
\end{equation}
and $0\le\bar{\lambda}\le1$. 
The convex quasi-linearity of some classes of quantum operations 
was noted by Kraus in
\cite{Kraus1983} (although Kraus didn't use this name).
In fact, the applicability of the selective 
measurements to the quantum mechanics is strongly related to the 
quasi-linearity property of this stochastic operation. 
Indeed, it implies that the set of ensembles representing a fixed  density matrix $\rho$ 
is mapped on the set of ensembles representing $\Phi(\rho)$.

\textit{The nonlinear evolution satisfying the quasi-linearity condition.}---In view 
of the above discussion it seems that there are no objections to consider
deterministic convex quasi-linear evolutions. Such an 
evolution map $\rho(t)=\Phi_t[\rho_0]$, with the initial condition 
$\rho(0)=\rho_0$ should form a semigroup satisfying  the relation (\ref{quasi_linearity})
for each value of the time parameter $t$; namely if
\begin{equation}
\lambda \rho_{a0} + (1-\lambda) \rho_{b0} = \rho_0
\label{rho0_ansamble}
\end{equation}
then
\begin{multline}
\Phi_t[\lambda \rho_{a0} + (1-\lambda) \rho_{b0}]\\
= \lambda(t) \Phi_t[\rho_{a0}] + (1-\lambda(t)) \Phi_t[\rho_{b0}],
\label{quasi_linearity_time_dep}
\end{multline}
with the conditions $\lambda(0)=\lambda$ and $0\le\lambda(t)\le1$.
To show that the set of deterministic convex quasi-linear evolutions
is non-empty we will construct a simple toy model of a qubit
evolution satisfying (\ref{quasi_linearity_time_dep}).

To define the evolution of a qubit we should determine nonlinear evolution 
of its Bloch vector. 
In construction of our model we use the well known
transformation rule for a three-velocity $\vec{v}$
under Lorentz boosts in a given direction, say $\vec{e}$ ($|\vec{e}|=1$):
\begin{equation}
\vec{v}^\prime = 
\frac{\vec{v}+\vec{e}[\sinh\eta+(\cosh\eta-1)(\vec{e}\cdot\vec{v})]}%
{\cosh\eta+(\vec{e}\cdot\vec{v})\sinh\eta},
\label{velocity_transf}
\end{equation}
where $\eta$ is the rapidity (we work in natural units with $\hbar=c=1$). 
The above transformation (\ref{velocity_transf}) is one of standard examples
of nonlinear transformations appearing in physics.

Let us notice two obvious facts:
(i) Boosts in a given direction form a one-parameter subgroup 
of the Lorentz group,
(ii) Length of a three-velocity is always less than or equal to 1.

Observation (i) allows us to treat the transformation rule 
(\ref{velocity_transf}) as
an equation defining time evolution of a three vector. Namely, we can rewrite
Eq.~(\ref{velocity_transf}) as 
\begin{equation}
\vec{n}(t)  = 
\frac{\vecg{\xi}+\vec{e}[\sinh(gt)+(\cosh(gt)-1)(\vec{e}\cdot\vecg{\xi})]}%
{\cosh(gt)+(\vec{e}\cdot\vecg{\xi})\sinh(gt)},
\label{evolution_xi}
\end{equation}
where the constant $g$ has been introduced from dimensional reasons and
$\vec{n}(0)\equiv\vecg{\xi}$ and $\vecg{\xi}^2\le1$. 
Now, 
point (i) implies that if we write $\vec{n}(t)=f_t(\vecg{\xi})$ then
$f_{t_2}\circ f_{t_1} = f_{t_1 + t_2}$,
therefore, (\ref{evolution_xi}) is a nonlinear time evolution of $\vec{n}(t)$.

Next, point (ii) gives us the possibility of identifying the vector $\vec{n}(t)$
with the Bloch vector defining a qubit density matrix $\rho(t)$:
\begin{equation}
\rho(t) = \tfrac{1}{2}\big( I + \vec{n}(t)\cdot\vecg{\sigma} \big).
\label{qubit_t}
\end{equation}
It means that the evolution $f_t$ corresponds to a nonlinear evolution of a qubit
density matrix:
\begin{equation}
\rho(t) = \Phi_t[\rho_0],
\end{equation}
where $\rho_0 = \tfrac{1}{2}\big( I + \vecg{\xi}\cdot\vecg{\sigma} \big)$.
Point (ii) implies that the condition
 $\vec{n}(t)^2\le1$ is preserved for all $t$.
Because the magnitude of a unit vector does not change under
the evolution (\ref{evolution_xi})
the subset of pure states is invariant under the
evolution $\Phi_t$.

Moreover, we can easily check that $\vec{n}(t)\to\vec{e}$ in the limit 
$t\to\infty$. Therefore, under the evolution (\ref{evolution_xi}) mixed
states evolve into pure states. In such cases the von Neumann entropy
of the state (\ref{qubit_t}) decreases during the evolution.
This observation, together with the second law of thermodynamics,
suggest that a carrier physical system (e.g. a spin one-half particle
or a two level atom) of the
qubit evolving according to Eq.~(\ref{evolution_xi})
cannot be isolated. Instead, it should be treated as an open quantum
system interacting with an environment.
Of course, it does not exclude the use of the evolution (\ref{evolution_xi})
in the considered Gisin Gedanken experiment.

Let us notice that the Bloch vector (\ref{evolution_xi}) is a solution
of the following nonlinear differential equation:
\begin{equation}
\dot{\vec{n}} = g 
\big( 
\vec{e} - \vec{n} (\vec{e}\cdot\vec{n}) 
\big)
\label{diff_equation}
\end{equation}
under the initial condition $\vec{n}(0)=\vecg{\xi}$.

Now, using the evolution (\ref{evolution_xi}) we can show that if the 
Eq.~(\ref{rho0_ansamble}) holds then
\begin{equation}
\rho(t) = \lambda(t) \rho_a(t) + (1-\lambda(t)) \rho_b(t),
\label{rhot_ansamble}
\end{equation}
where
\begin{equation}
\rho_a(t) = \tfrac{1}{2}\big( I + \vec{n}_a(t)\cdot\vecg{\sigma} \big),\quad
\rho_b(t) = \tfrac{1}{2}\big( I + \vec{n}_b(t)\cdot\vecg{\sigma} \big)
\label{initial_rho_a_b}
\end{equation}
and both Bloch vectors $\vec{n}_a$ and $\vec{n}_b$
evolve according to the nonlinear law (\ref{evolution_xi}) under the replacement
$\vecg{\xi}\to\vecg{\xi}_{a}$ or $\vecg{\xi}_{b}$, respectively.
Using Eqs.~(\ref{initial_rho_a_b},\ref{rhot_ansamble},\ref{evolution_xi})
we can find the coefficient $\lambda(t)$. It is given by
\begin{equation}
\lambda(t) = \frac{1 + (\vec{e}\cdot\vecg{\xi}_{a})\tanh(gt)}%
{1 + (\vec{e}\cdot\vecg{\xi})\tanh(gt)} \,\,
\lambda.
\label{lambda_t}
\end{equation}
Therefore, we can conclude that each ensemble (\ref{rho0_ansamble}) equivalent to
$\rho_0$ evolves under prescription (\ref{evolution_xi}) into ensemble 
(\ref{rhot_ansamble}) equivalent to $\rho(t)$.
We see that the coefficient $\lambda(t)$ explicitly depends on time.
However, in view of our previous remark that qubit evolving
according to Eq.~(\ref{evolution_xi}) cannot be treated as an isolated 
system, the dependence of $\lambda$ on time is rather expected.

Returning to the Gisin Gedanken experiment, the ensemble
$\rho_B=\tr_A(\rho_\varphi)$, accessible to the observer 
B [Eq.~(\ref{rho_B})],
evolves under (\ref{evolution_xi}) as follows
\begin{equation}
\rho_B(t) = 
\tfrac{1}{2} \big( I + (\vec{e}\cdot\vecg{\sigma}) \tanh(gt) \big),
\end{equation}
and
\begin{multline}
\rho_{B_{\varphi}^{1}}(t)  = \tfrac{1}{2} \bigg(  
I \\
+ \frac{-\vecg{\zeta}_\varphi  + 
\big[ 
\sinh(gt) - 
(\vec{e} \cdot \vecg{\zeta}_\varphi)
\big( \cosh(gt) -1 \big) 
\big]\vec{e}}{\cosh(gt) 
- (\vec{e}\cdot\vecg{\zeta}_\varphi)\sinh(gt)}
\cdot \vecg{\sigma}
\bigg),
\end{multline}
\begin{multline}
\rho_{B_{\varphi}^{2}}(t)  = \tfrac{1}{2} \bigg(  
I \\
+ \frac{\vecg{\zeta}_\varphi  + 
	\big[ 
	\sinh(gt) + 
	(\vec{e} \cdot \vecg{\zeta}_\varphi)
	\big( \cosh(gt) -1 \big) 
	\big] \vec{e} }{\cosh(gt) 
	+ (\vec{e}\cdot\vecg{\zeta}_\varphi)\sinh(gt)}
\cdot \vecg{\sigma}
\bigg).
\end{multline}
where $\vecg{\zeta}_\varphi$ is given in Eq.~(\ref{zeta}).
So using (\ref{lambda_t}) we find that in this case
\begin{equation}
\lambda(t) = \tfrac{1}{2} 
\big( 
1-(\vec{e}\cdot\vecg{\zeta}_\varphi)\tanh(gt) 
\big).
\end{equation}

Therefore, finally, it really holds that:
\begin{multline}
\lambda(t) \rho_{B_{\varphi}^{1}}(t) 
+ (1-\lambda(t)) \rho_{B_{\varphi}^{2}}(t)\\
 = \tfrac{1}{2} \big( I + (\vec{e}\cdot\vecg{\sigma}) \tanh(gt) \big)
 = \rho_B(t).
\end{multline}
Thus, observer B cannot register any change of the polarization by 
the observer A, exactly as in the standard case.

\textit{Conclusions.}---We have shown that time evolutions
satisfying the quasi-linearity property (\ref{quasi_linearity_time_dep})
are admissible
in the convex set of density operators even if they are nonlinear.
As an example we discussed nonlinear time evolution of a qubit explicitly
satisfying this property and we applied it to the famous Gedanken
nonlocal correlation experiment by Gisin \cite{Gisin1990_Weinberg_nonlin}.
We concluded that this evolution does not 
allow for arbitrary fast signaling.
Therefore, such an evolution is not in contradiction with
the special relativity at this level.
It remains an open question how big is the class of convex quasi-linear
evolutions,
work on this subject is in progress \cite{RC2019_in_prep}. 

We can notice that the differential equation for the Bloch vector
(\ref{diff_equation}) resembles the equation for the Bloch vector 
in the simplified Weinberg model, 
discussed in \cite{Gisin1990_Weinberg_nonlin}, adapted to the qubit case:
\begin{equation}
\dot{\vec{n}} = 
g \big(
\vec{e}\times\vec{n}(\vec{e}\cdot\vec{n})
\big).
\label{Weinberg_evol}
\end{equation}
Solution of the last equation reads
\begin{multline}
\vec{n}(t) = 
\vecg{\xi} \cos\theta(t)
+ (\vec{e}\times\vecg{\xi}) \sin\theta(t)\\
+ \vec{e} (\vec{e}\cdot\vecg{\xi})
\big(1-\cos\theta(t)\big),
\label{Weinberg_evol_sol}
\end{multline}
where $\theta(t) = g t (\vec{e}\cdot\vecg{\xi})$ and 
$\vec{n}(0)=\vecg{\xi}$.
However, it can be explicitly shown that in general solutions 
(\ref{Weinberg_evol_sol}) cannot satisfy the quasi-linearity
property (\ref{rhot_ansamble}).
Consequently, evolution in the Weinberg model allows for arbitrary 
fast signaling.

\begin{acknowledgments}
We are grateful to Marek Czachor and Krzysztof Kowalski for 
interesting discussion.
This work has been supported by the Polish National Science Centre
under the contract 2014/15/B/ST2/00117 and by the University of Lodz.
\end{acknowledgments}


\begin{thebibliography}{25}%
	\makeatletter
	\providecommand \@ifxundefined [1]{%
		\@ifx{#1\undefined}
	}%
	\providecommand \@ifnum [1]{%
		\ifnum #1\expandafter \@firstoftwo
		\else \expandafter \@secondoftwo
		\fi
	}%
	\providecommand \@ifx [1]{%
		\ifx #1\expandafter \@firstoftwo
		\else \expandafter \@secondoftwo
		\fi
	}%
	\providecommand \natexlab [1]{#1}%
	\providecommand \enquote  [1]{``#1''}%
	\providecommand \bibnamefont  [1]{#1}%
	\providecommand \bibfnamefont [1]{#1}%
	\providecommand \citenamefont [1]{#1}%
	\providecommand \href@noop [0]{\@secondoftwo}%
	\providecommand \href [0]{\begingroup \@sanitize@url \@href}%
	\providecommand \@href[1]{\@@startlink{#1}\@@href}%
	\providecommand \@@href[1]{\endgroup#1\@@endlink}%
	\providecommand \@sanitize@url [0]{\catcode `\\12\catcode `\$12\catcode
		`\&12\catcode `\#12\catcode `\^12\catcode `\_12\catcode `\%12\relax}%
	\providecommand \@@startlink[1]{}%
	\providecommand \@@endlink[0]{}%
	\providecommand \url  [0]{\begingroup\@sanitize@url \@url }%
	\providecommand \@url [1]{\endgroup\@href {#1}{\urlprefix }}%
	\providecommand \urlprefix  [0]{URL }%
	\providecommand \Eprint [0]{\href }%
	\providecommand \doibase [0]{https://doi.org/}%
	\providecommand \selectlanguage [0]{\@gobble}%
	\providecommand \bibinfo  [0]{\@secondoftwo}%
	\providecommand \bibfield  [0]{\@secondoftwo}%
	\providecommand \translation [1]{[#1]}%
	\providecommand \BibitemOpen [0]{}%
	\providecommand \bibitemStop [0]{}%
	\providecommand \bibitemNoStop [0]{.\EOS\space}%
	\providecommand \EOS [0]{\spacefactor3000\relax}%
	\providecommand \BibitemShut  [1]{\csname bibitem#1\endcsname}%
	\let\auto@bib@innerbib\@empty
	\bibitem [{\citenamefont {Bia{\l}ynicki-Birula}\ and\ \citenamefont
		{Mycielski}(1976)}]{BM1976_Nonlinear}%
	\BibitemOpen
	\bibfield  {author} {\bibinfo {author} {\bibfnamefont {I.}~\bibnamefont
			{Bia{\l}ynicki-Birula}}\ and\ \bibinfo {author} {\bibfnamefont
			{J.}~\bibnamefont {Mycielski}},\ }\bibfield  {title} {\bibinfo {title}
		{Nonlinear wave mechanics},\ }\href@noop {} {\bibfield  {journal} {\bibinfo
			{journal} {Ann. Phys. (New York)}\ }\textbf {\bibinfo {volume} {100}},\
		\bibinfo {pages} {62} (\bibinfo {year} {1976})}\BibitemShut {NoStop}%
	\bibitem [{\citenamefont
		{Weinberg}(1989{\natexlab{a}})}]{Weinberg1989_Nonlin_Annals}%
	\BibitemOpen
	\bibfield  {author} {\bibinfo {author} {\bibfnamefont {S.}~\bibnamefont
			{Weinberg}},\ }\bibfield  {title} {\bibinfo {title} {Testing quantum
			mechanics},\ }\href
	{https://doi.org/https://doi.org/10.1016/0003-4916(89)90276-5} {\bibfield
		{journal} {\bibinfo  {journal} {Ann. Phys. (New York)}\ }\textbf {\bibinfo
			{volume} {194}},\ \bibinfo {pages} {336 } (\bibinfo {year}
		{1989}{\natexlab{a}})}\BibitemShut {NoStop}%
	\bibitem [{\citenamefont
		{Weinberg}(1989{\natexlab{b}})}]{Weinberg1989_PRL_nonlin}%
	\BibitemOpen
	\bibfield  {author} {\bibinfo {author} {\bibfnamefont {S.}~\bibnamefont
			{Weinberg}},\ }\bibfield  {title} {\bibinfo {title} {Precision tests of
			quantum mechanics},\ }\href@noop {} {\bibfield  {journal} {\bibinfo
			{journal} {Phys. Rev. Lett.}\ }\textbf {\bibinfo {volume} {62}},\ \bibinfo
		{pages} {485} (\bibinfo {year} {1989}{\natexlab{b}})}\BibitemShut {NoStop}%
	\bibitem [{\citenamefont {Doebner}\ and\ \citenamefont
		{Goldin}(1992)}]{DG1992_Nonlin_PLA}%
	\BibitemOpen
	\bibfield  {author} {\bibinfo {author} {\bibfnamefont {H.-D.}\ \bibnamefont
			{Doebner}}\ and\ \bibinfo {author} {\bibfnamefont {G.~A.}\ \bibnamefont
			{Goldin}},\ }\bibfield  {title} {\bibinfo {title} {On a general nonlinear
			{S}chr{\"o}dinger equation admitting diffusion currents},\ }\href
	{https://doi.org/https://doi.org/10.1016/0375-9601(92)90061-P} {\bibfield
		{journal} {\bibinfo  {journal} {Phys. Lett. A}\ }\textbf {\bibinfo {volume}
			{162}},\ \bibinfo {pages} {397 } (\bibinfo {year} {1992})}\BibitemShut
	{NoStop}%
	\bibitem [{\citenamefont {Doebner}\ and\ \citenamefont
		{Goldin}(1996)}]{DG1996_Nonlin_PhysRevA.54.3764}%
	\BibitemOpen
	\bibfield  {author} {\bibinfo {author} {\bibfnamefont {H.-D.}\ \bibnamefont
			{Doebner}}\ and\ \bibinfo {author} {\bibfnamefont {G.~A.}\ \bibnamefont
			{Goldin}},\ }\bibfield  {title} {\bibinfo {title} {Introducing nonlinear
			gauge transformations in a family of nonlinear {S}chr\"odinger equations},\
	}\href {https://doi.org/10.1103/PhysRevA.54.3764} {\bibfield  {journal}
		{\bibinfo  {journal} {Phys. Rev. A}\ }\textbf {\bibinfo {volume} {54}},\
		\bibinfo {pages} {3764} (\bibinfo {year} {1996})}\BibitemShut {NoStop}%
	\bibitem [{\citenamefont {Gisin}\ and\ \citenamefont
		{Rigo}(1995)}]{GR1995_relevant_Sch}%
	\BibitemOpen
	\bibfield  {author} {\bibinfo {author} {\bibfnamefont {N.}~\bibnamefont
			{Gisin}}\ and\ \bibinfo {author} {\bibfnamefont {M.}~\bibnamefont {Rigo}},\
	}\bibfield  {title} {\bibinfo {title} {Relevant and irrelevant nonlinear
			{S}chr\"odinger equations},\ }\href@noop {} {\bibfield  {journal} {\bibinfo
			{journal} {J. Phys. A: Math. Gen.}\ }\textbf {\bibinfo {volume} {28}},\
		\bibinfo {pages} {7375} (\bibinfo {year} {1995})}\BibitemShut {NoStop}%
	\bibitem [{\citenamefont {Gisin}(1990)}]{Gisin1990_Weinberg_nonlin}%
	\BibitemOpen
	\bibfield  {author} {\bibinfo {author} {\bibfnamefont {N.}~\bibnamefont
			{Gisin}},\ }\bibfield  {title} {\bibinfo {title} {Weinberg's non-linear
			quantum mechanics and supraluminal communications},\ }\href@noop {}
	{\bibfield  {journal} {\bibinfo  {journal} {Phys. Lett. A}\ }\textbf
		{\bibinfo {volume} {143}},\ \bibinfo {pages} {1} (\bibinfo {year}
		{1990})}\BibitemShut {NoStop}%
	\bibitem [{\citenamefont {Polchinski}(1991)}]{Polchinski1991_on_Weinberg}%
	\BibitemOpen
	\bibfield  {author} {\bibinfo {author} {\bibfnamefont {J.}~\bibnamefont
			{Polchinski}},\ }\bibfield  {title} {\bibinfo {title} {Weinberg's nonlinear
			quantum mechanics and the {E}instein-{P}odolsky-{R}osen paradox},\ }\href
	{https://doi.org/10.1103/PhysRevLett.66.397} {\bibfield  {journal} {\bibinfo
			{journal} {Phys. Rev. Lett.}\ }\textbf {\bibinfo {volume} {66}},\ \bibinfo
		{pages} {397} (\bibinfo {year} {1991})}\BibitemShut {NoStop}%
	\bibitem [{\citenamefont {Czachor}(1991)}]{Czachor1991}%
	\BibitemOpen
	\bibfield  {author} {\bibinfo {author} {\bibfnamefont {M.}~\bibnamefont
			{Czachor}},\ }\bibfield  {title} {\bibinfo {title} {Mobility and
			non-separability},\ }\href {https://doi.org/10.1007/BF00665894} {\bibfield
		{journal} {\bibinfo  {journal} {Found. Phys. Lett.}\ }\textbf {\bibinfo
			{volume} {4}},\ \bibinfo {pages} {351} (\bibinfo {year} {1991})}\BibitemShut
	{NoStop}%
	\bibitem [{\citenamefont {Czachor}(1998)}]{Czachor1998_nonlocal-eq-NLQM}%
	\BibitemOpen
	\bibfield  {author} {\bibinfo {author} {\bibfnamefont {M.}~\bibnamefont
			{Czachor}},\ }\bibfield  {title} {\bibinfo {title} {Nonlocal-looking
			equations can make nonlinear quantum dynamics local},\ }\href
	{https://doi.org/10.1103/PhysRevA.57.4122} {\bibfield  {journal} {\bibinfo
			{journal} {Phys. Rev. A}\ }\textbf {\bibinfo {volume} {57}},\ \bibinfo
		{pages} {4122} (\bibinfo {year} {1998})}\BibitemShut {NoStop}%
	\bibitem [{\citenamefont {Czachor}\ and\ \citenamefont
		{Doebner}(2002)}]{CD2002_Correl-exp-nonlin_PLA}%
	\BibitemOpen
	\bibfield  {author} {\bibinfo {author} {\bibfnamefont {M.}~\bibnamefont
			{Czachor}}\ and\ \bibinfo {author} {\bibfnamefont {H.-D.}\ \bibnamefont
			{Doebner}},\ }\bibfield  {title} {\bibinfo {title} {Correlation experiments
			in nonlinear quantum mechanics},\ }\href
	{https://doi.org/https://doi.org/10.1016/S0375-9601(02)00959-3} {\bibfield
		{journal} {\bibinfo  {journal} {Phys. Lett. A}\ }\textbf {\bibinfo {volume}
			{301}},\ \bibinfo {pages} {139 } (\bibinfo {year} {2002})}\BibitemShut
	{NoStop}%
	\bibitem [{\citenamefont {Kent}(2005)}]{Kent2005_Nonlin_superluminality}%
	\BibitemOpen
	\bibfield  {author} {\bibinfo {author} {\bibfnamefont {A.}~\bibnamefont
			{Kent}},\ }\bibfield  {title} {\bibinfo {title} {Nonlinearity without
			superluminality},\ }\href {https://doi.org/10.1103/PhysRevA.72.012108}
	{\bibfield  {journal} {\bibinfo  {journal} {Phys. Rev. A}\ }\textbf {\bibinfo
			{volume} {72}},\ \bibinfo {pages} {012108} (\bibinfo {year}
		{2005})}\BibitemShut {NoStop}%
	\bibitem [{\citenamefont {Ferrero}\ \emph {et~al.}(2004)\citenamefont
		{Ferrero}, \citenamefont {Salgado},\ and\ \citenamefont
		{S\'anchez-G\'omez}}]{FSSG2004_NLQM-not-supraluminal}%
	\BibitemOpen
	\bibfield  {author} {\bibinfo {author} {\bibfnamefont {M.}~\bibnamefont
			{Ferrero}}, \bibinfo {author} {\bibfnamefont {D.}~\bibnamefont {Salgado}},\
		and\ \bibinfo {author} {\bibfnamefont {J.~L.}\ \bibnamefont
			{S\'anchez-G\'omez}},\ }\bibfield  {title} {\bibinfo {title} {Nonlinear
			quantum evolution does not imply supraluminal communication},\ }\href
	{https://doi.org/10.1103/PhysRevA.70.014101} {\bibfield  {journal} {\bibinfo
			{journal} {Phys. Rev. A}\ }\textbf {\bibinfo {volume} {70}},\ \bibinfo
		{pages} {014101} (\bibinfo {year} {2004})}\BibitemShut {NoStop}%
	\bibitem [{\citenamefont {Jordan}(2009)}]{Jordan2009_Why-q-dyn-is-linear}%
	\BibitemOpen
	\bibfield  {author} {\bibinfo {author} {\bibfnamefont {T.~F.}\ \bibnamefont
			{Jordan}},\ }\bibfield  {title} {\bibinfo {title} {Why quantum dynamics is
			linear},\ }\href {https://doi.org/10.1088/1742-6596/196/1/012010} {\bibfield
		{journal} {\bibinfo  {journal} {Journal of Physics: Conference Series}\
		}\textbf {\bibinfo {volume} {196}},\ \bibinfo {pages} {012010} (\bibinfo
		{year} {2009})}\BibitemShut {NoStop}%
	\bibitem [{\citenamefont {Helou}\ and\ \citenamefont
		{Chen}(2017)}]{HCh2017_Born_NLQM}%
	\BibitemOpen
	\bibfield  {author} {\bibinfo {author} {\bibfnamefont {B.}~\bibnamefont
			{Helou}}\ and\ \bibinfo {author} {\bibfnamefont {Y.}~\bibnamefont {Chen}},\
	}\bibfield  {title} {\bibinfo {title} {Extensions of {B}orn's rule to
			non-linear quantum mechanics, some of which do not imply superluminal
			communication},\ }\href {https://doi.org/10.1088/1742-6596/880/1/012021}
	{\bibfield  {journal} {\bibinfo  {journal} {Journal of Physics: Conference
				Series}\ }\textbf {\bibinfo {volume} {880}},\ \bibinfo {pages} {012021}
		(\bibinfo {year} {2017})}\BibitemShut {NoStop}%
	\bibitem [{\citenamefont {Mielnik}(2001)}]{Mielnik2001_NLQM}%
	\BibitemOpen
	\bibfield  {author} {\bibinfo {author} {\bibfnamefont {B.}~\bibnamefont
			{Mielnik}},\ }\bibfield  {title} {\bibinfo {title} {Nonlinear quantum
			mechanics: a conflict with the ptolomean structure?},\ }\href
	{https://doi.org/https://doi.org/10.1016/S0375-9601(01)00583-7} {\bibfield
		{journal} {\bibinfo  {journal} {Phys. Lett. A}\ }\textbf {\bibinfo {volume}
			{289}},\ \bibinfo {pages} {1 } (\bibinfo {year} {2001})}\BibitemShut
	{NoStop}%
	\bibitem [{\citenamefont {Jordan}(2010)}]{Jordan2010_Test_NLQM}%
	\BibitemOpen
	\bibfield  {author} {\bibinfo {author} {\bibfnamefont {T.~F.}\ \bibnamefont
			{Jordan}},\ }\bibfield  {title} {\bibinfo {title} {Fundamental significance
			of tests that quantum dynamics is linear},\ }\href
	{https://doi.org/10.1103/PhysRevA.82.032103} {\bibfield  {journal} {\bibinfo
			{journal} {Phys. Rev. A}\ }\textbf {\bibinfo {volume} {82}},\ \bibinfo
		{pages} {032103} (\bibinfo {year} {2010})}\BibitemShut {NoStop}%
	\bibitem [{\citenamefont {Ghirardi}\ \emph {et~al.}(1986)\citenamefont
		{Ghirardi}, \citenamefont {Rimini},\ and\ \citenamefont {Weber}}]{GRW1986}%
	\BibitemOpen
	\bibfield  {author} {\bibinfo {author} {\bibfnamefont {G.~C.}\ \bibnamefont
			{Ghirardi}}, \bibinfo {author} {\bibfnamefont {A.}~\bibnamefont {Rimini}},\
		and\ \bibinfo {author} {\bibfnamefont {T.}~\bibnamefont {Weber}},\ }\bibfield
	{title} {\bibinfo {title} {Unified dynamics for microscopic and macroscopic
			systems},\ }\href {https://doi.org/10.1103/PhysRevD.34.470} {\bibfield
		{journal} {\bibinfo  {journal} {Phys. Rev. D}\ }\textbf {\bibinfo {volume}
			{34}},\ \bibinfo {pages} {470} (\bibinfo {year} {1986})}\BibitemShut
	{NoStop}%
	\bibitem [{\citenamefont {Bassi}\ \emph {et~al.}(2013)\citenamefont {Bassi},
		\citenamefont {Lochan}, \citenamefont {Satin}, \citenamefont {Singh},\ and\
		\citenamefont {Ulbricht}}]{BLSSU2013}%
	\BibitemOpen
	\bibfield  {author} {\bibinfo {author} {\bibfnamefont {A.}~\bibnamefont
			{Bassi}}, \bibinfo {author} {\bibfnamefont {K.}~\bibnamefont {Lochan}},
		\bibinfo {author} {\bibfnamefont {S.}~\bibnamefont {Satin}}, \bibinfo
		{author} {\bibfnamefont {T.~P.}\ \bibnamefont {Singh}},\ and\ \bibinfo
		{author} {\bibfnamefont {H.}~\bibnamefont {Ulbricht}},\ }\bibfield  {title}
	{\bibinfo {title} {Models of wave-function collapse, underlying theories, and
			experimental tests},\ }\href@noop {} {\bibfield  {journal} {\bibinfo
			{journal} {Rev. Mod. Phys.}\ }\textbf {\bibinfo {volume} {85}},\ \bibinfo
		{pages} {471} (\bibinfo {year} {2013})}\BibitemShut {NoStop}%
	\bibitem [{\citenamefont {Gisin}(1989)}]{Gisin1989_stochastic_q_dyn_relativ}%
	\BibitemOpen
	\bibfield  {author} {\bibinfo {author} {\bibfnamefont {N.}~\bibnamefont
			{Gisin}},\ }\bibfield  {title} {\bibinfo {title} {Stochastic quantum dynamics
			and relaticity},\ }\href@noop {} {\bibfield  {journal} {\bibinfo  {journal}
			{Helv. Phys. Acta}\ }\textbf {\bibinfo {volume} {62}},\ \bibinfo {pages}
		{363} (\bibinfo {year} {1989})}\BibitemShut {NoStop}%
	\bibitem [{\citenamefont {Jordan}(1991)}]{Jordan1991_assump-Schroedinger}%
	\BibitemOpen
	\bibfield  {author} {\bibinfo {author} {\bibfnamefont {T.~F.}\ \bibnamefont
			{Jordan}},\ }\bibfield  {title} {\bibinfo {title} {Assumptions implying the
			{S}chrödinger equation},\ }\href {https://doi.org/10.1119/1.16780}
	{\bibfield  {journal} {\bibinfo  {journal} {Am. J. Phys.}\ }\textbf {\bibinfo
			{volume} {59}},\ \bibinfo {pages} {606} (\bibinfo {year} {1991})}\BibitemShut
	{NoStop}%
	\bibitem [{\citenamefont
		{Jordan}(2006)}]{Jordan2006_PhysRevA_Assump-imply-linear}%
	\BibitemOpen
	\bibfield  {author} {\bibinfo {author} {\bibfnamefont {T.~F.}\ \bibnamefont
			{Jordan}},\ }\bibfield  {title} {\bibinfo {title} {Assumptions that imply
			quantum dynamics is linear},\ }\href
	{https://doi.org/10.1103/PhysRevA.73.022101} {\bibfield  {journal} {\bibinfo
			{journal} {Phys. Rev. A}\ }\textbf {\bibinfo {volume} {73}},\ \bibinfo
		{pages} {022101} (\bibinfo {year} {2006})}\BibitemShut {NoStop}%
	\bibitem [{\citenamefont {Bassi}\ and\ \citenamefont
		{Hejazi}(2015)}]{BK2015_Faster-then-light-linear}%
	\BibitemOpen
	\bibfield  {author} {\bibinfo {author} {\bibfnamefont {A.}~\bibnamefont
			{Bassi}}\ and\ \bibinfo {author} {\bibfnamefont {K.}~\bibnamefont {Hejazi}},\
	}\bibfield  {title} {\bibinfo {title} {No-faster-than-light-signaling implies
			linear evolution. {A} re-derivation},\ }\href
	{https://doi.org/10.1088/0143-0807/36/5/055027} {\bibfield  {journal}
		{\bibinfo  {journal} {European J. Phys.}\ }\textbf {\bibinfo {volume} {36}},\
		\bibinfo {pages} {055027} (\bibinfo {year} {2015})}\BibitemShut {NoStop}%
	\bibitem [{\citenamefont {Kraus}(1983)}]{Kraus1983}%
	\BibitemOpen
	\bibfield  {author} {\bibinfo {author} {\bibfnamefont {K.}~\bibnamefont
			{Kraus}},\ }\href@noop {} {\emph {\bibinfo {title} {States, Effects, and
				Operations}}}\ (\bibinfo  {publisher} {Springer-Verlag},\ \bibinfo {address}
	{Berlin, Heidelberg, New York, Tokyo},\ \bibinfo {year} {1983})\BibitemShut
	{NoStop}%
	\bibitem [{\citenamefont {Rembieli\'nski}\ and\ \citenamefont
		{Caban}()}]{RC2019_in_prep}%
	\BibitemOpen
	\bibfield  {author} {\bibinfo {author} {\bibfnamefont {J.}~\bibnamefont
			{Rembieli\'nski}}\ and\ \bibinfo {author} {\bibfnamefont {P.}~\bibnamefont
			{Caban}},\ }\href@noop {} {\ }\bibinfo {note} {In preparation}\BibitemShut
	{NoStop}%
\end{thebibliography}

%

\end{document}